\documentclass{emulateapj}
\newcommand{\km}{${\rm km\,s}^{-1}$}
\newcommand{\fuse}{{\em FUSE}}
\newcommand{\hi}{H$\;${\small\rm I}\relax}

\newcommand{\cii}{C$\;${\small\rm II}\relax}

\newcommand{\ciii}{C$\;${\small\rm III}\relax}
\newcommand{\civ}{C$\;${\small\rm IV}\relax}
\newcommand{\nni}{N$\;${\small\rm I}\relax}

\newcommand{\nv}{N$\;${\small\rm V}\relax}
\newcommand{\oi}{O$\;${\small\rm I}\relax}

\newcommand{\ovi}{O$\;${\small\rm VI}\relax}

\newcommand{\sii}{S$\;${\small\rm II}\relax}
\newcommand{\siii}{Si$\;${\small\rm II}\relax}
\newcommand{\siiii}{Si$\;${\small\rm III}\relax}

\newcommand{\alii}{Al$\;${\small\rm II}\relax}
\newcommand{\Siii}{S$\;${\small\rm III}\relax}
\newcommand{\siiv}{Si$\;${\small\rm IV}\relax}

\newcommand{\feii}{Fe$\;${\small\rm II}\relax}

\newcommand{\Niii}{Ni$\;${\small\rm II}\relax}

\newcommand{\vlsr}{$v_{\rm LSR}$\relax}
\newcommand{\hst}{{\em HST}}

\slugcomment{Submitted to the ApJ letters}
\shortauthors{Lehner \& Howk}
\shorttitle{Distance of  the Highly Ionized HVCs}
\begin{document}

\title{Origin(s) of  the Highly Ionized High-Velocity Clouds Based on Their Distances\altaffilmark{1}}
\author{N.\ Lehner and J.C. \ Howk
	}
\affil{Department of Physics, University of Notre Dame, 225 Nieuwland Science Hall, Notre Dame, IN 46556}

\altaffiltext{1}{Based on observations made with the NASA/ESA Hubble Space Telescope,
obtained at the Space Telescope Science Institute, which is operated by the
Association of Universities for Research in Astronomy, Inc. under NASA
contract No. NAS5-26555.}

\begin{abstract}
Previous \hst\ and \fuse\ observations have revealed highly ionized high-velocity clouds (HVCs) or more generally low \hi\ column  HVCs along extragalactic sightlines over 70--90\% of the sky. The distances of these HVCs have remained largely unknown hampering to distinguish  a ``Galactic" origin (e.g., outflow, inflow) from a ``Local Group" origin (e.g., warm-hot intergalactic medium). We present the first detection of highly ionized HVCs in the Cosmic Origins Spectrograph (COS) spectrum of the early-type star HS\,1914+7134 ($l = 103\degr$, $b=+24 \degr$) located in the outer region of the Galaxy at $d \simeq 14.9$ kpc. Two HVCs are detected in absorption at $v_{\rm LSR} = -118$ and $-180$ \km\ in several species, including \civ, \siiv, \siiii, \alii, \cii, \siii, \oi, but \hi\ 21-cm emission is only seen at $-118$ \km. Within 17\degr\ of  HS\,1914+7134, we found HVC absorption of low and high ions at similar velocities toward 5 extragalactic sightlines, suggesting that these HVCs are related. The component at $-118$ \km\ is likely associated with the Outer Arm of the Milky Way. The highly ionized HVC at $-180$ \km\  is an HVC plunging at high speed onto the thick disk of the Milky Way. This is the second detection of highly ionized HVCs toward Galactic stars, supporting a ``Galactic" origin for at least some of these clouds. 
\end{abstract}

\section{Introduction}
The Milky Way is surrounded by large gaseous complexes observed at high velocities ($|\mbox{\vlsr}| \ga 90$ \km), known as HVCs. HVCs were traditionally regarded as neutral entities \citep[e.g.,][]{wakker97}, but this view has given way to one that recognizes the importance of an ionized component to many of these clouds thanks to high quality UV observations from {\hst} and \fuse\ \citep{sembach95,sembach99,sembach03,lehner01a,lehner02,collins04,collins05,collins09,ganguly05,fox04,fox05,fox06,richter09,shull09}. Many of these  HVCs are almost completely ionized and show  absorption from the ``high ions'' \ovi, \civ, and \siiv.  These are therefore deemed highly ionized HVCs, although they often show absorption  not only in the high ions, but also in the lower ions (e.g., \cii, \ciii) and even atoms (\hi, \oi).  A better designation would be multiphase HVCs or low \hi\ column HVCs \citep[see also][]{fox06,collins09}. The covering factor of these low \hi\ column density HVCs is extremely high and much larger than that of the \hi\ 21-cm HVCs. \fuse\ surveys of AGNs and QSOs show that 60--70\%\ (perhaps even 85\%) of the high-latitude sky is covered by \ovi\ HVCs, the majority of which ($\sim$75\%) have no \hi\ emission counterpart \citep{sembach03,fox06}. \citet{shull09} and \citet{collins09} show that high-velocity \siiii\ absorption has a detection rate of 80--90\%\ at high latitude. These numbers are to be contrasted to the \hi\ 21-cm emission ($N($\hi$)\ga 10^{17.9}$ cm$^{-2}$) covering factor of $\sim$37\% \citep{murphy95}.
 
Both galactic and larger-scale phenomena/structures have been invoked for the origin(s) of these HVCs, including galactic feedback processes, disk-halo mass exchange (e.g., through the accretion of matter condensing from an extended corona), or the shock-heated warm-hot intergalactic medium (WHIM) of the Local Group \citep[e.g.,][]{blitz99,sembach03,nicastro03,collins05,fox06}. Pros and cons of both theories have been discussed, but they are circumstantial and tentative, relying on assumptions of poorly-constrained quantities, such as metallicities and ionization conditions. As for the larger \hi\ column HVCs, the distance of the HVCs is the sole {\it direct} test for these competing models. 

Previous searches of  these types of HVCs in the spectra of Galactic halo stars have mostly failed \citep{zsargo03,sembach03}, leading some support to the hypothesis that the highly ionized HVCs could be mainly associated with the WHIM of the Local Group rather than with some galactic-scale phenomena. However, most distance limits for highly ionized HVCs are not very constraining, as the stars in the \citet{zsargo03} survey are mostly significantly closer than those against which the higher column density \hi\ HVCs have been detected, $5 \la d \la 15$ kpc \citep[e.g.,][]{wakker01,wakker08}. Further for the  stars at large $d$ in Zsarg\`o et al., the signal-to-noise (S/N) levels were often not high enough and/or the stellar continua were too complicated to be able to reliably detect \ovi\ at high velocity. 

In \citet{zech08}, we revisited one of the Zsarg\`o et al. stars, the PAGB star ZNG\,1 in the globular cluster NGC\,5904 ($z = 5.3$ kpc, see Table~\ref{t-sum}), with additional \fuse\ and new {\em HST}/STIS (E140M) observations, finding two highly ionized HVCs with no detected \hi\ emission. This firmly demonstrated that some of these HVCs originate near the Galactic disk. For sightlines passing through  known large \hi\ complexes (e.g., Complex C, Magellanic Stream), highly ionized gas is found with similar kinematics as  the neutral HVCs \citep[e.g.,][]{collins07,fox04}.  Kinematics and metallicity arguments  also suggest some of low \hi\ column HVCs observed toward the Large Magellanic Cloud (LMC) are due to galactic feedback occurring within the LMC \citep{lehner07,lehner09}.  

Progress in determining the distances of the highly ionized HVCs seen in the Galactic sky is now possible owing to the recent installation of the  Cosmic Origins Spectrograph (COS) (and repair of the Space Telescope Imaging Spectrograph, STIS) onboard of \hst. The high throughput of COS at medium resolution ($R\sim 20,000$) allows us to efficiently observe targets much fainter in the UV than previously, and hence observe more distant stars. In this Letter, we report on the first observations from our COS program to search and characterize the high velocity \nv, \civ, and \siiv\ interstellar absorption in stars at large distances  ($4<d<21$ kpc, $3<|z|<13$ kpc). Presently, four stars out of 24 have been observed. For two of them (PG1708+142, PG1704+222, PAGB stars at $d = 10$ and 7 kpc, respectively), no HVCs were detected. For another one the signal-to-noise was lower than expected and will need a closer inspection. However, in the COS spectrum of the fourth star (HS\,1914+7139, see Table~\ref{t-sum}), two HVCs at $v_{\rm LSR} = -180$ and $-118$ \km\ are detected in \civ, \siiv, \cii, and other species.  While the HVC at $-118$ \km\ is detected in both absorption and \hi\ 21-cm emission and may be associated with the Outer Arm of the Galaxy, the other HVC is only detected in absorption.  Thus, we have found another highly ionized HVC falling toward the sun and well within the Milky Way.

\section{Observations and Analysis}
As HVCs were only detected toward HS\,1914+7139, we only  summarize the data for this star. HS\,1914+7139 is a (possible runaway) B2.5\,IV star ($\log g = 3.9$, $T_{\rm eff} = 17,600$ K) with a projected rotational velocity $v\sin i = 250$ \km\ and a radial velocity in the LSR frame of $-25$ \km\ \citep{heber95,ramspeck01}. It is located in the outer region of the Galaxy at a distance from the sun $d  = 14.9$ kpc and height above the Galactic plane $z = 6.0$ kpc \citep{ramspeck01}.  The distance is accurate to $\sim$\,20\% and is based on  detailed model of the stellar atmosphere. The large projected rotational velocity is ideal for interstellar studies removing the possibility that some of the narrow lines observed in the spectrum are stellar. 

\begin{figure}[tbp]
\epsscale{0.5} 
\plotone{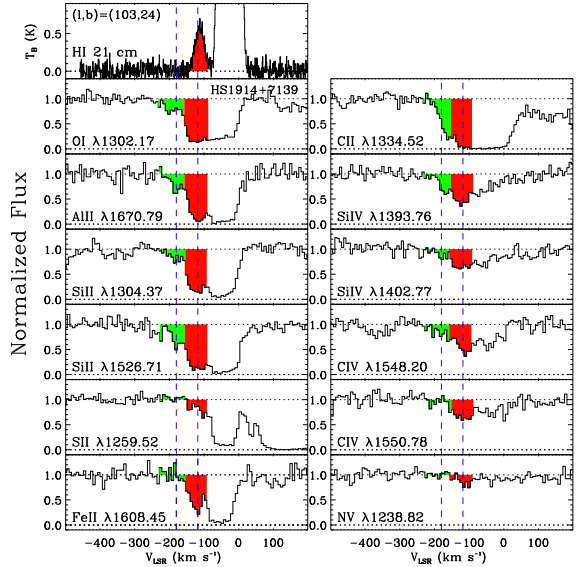}
\caption{Normalized profiles of some of the detected species observed in the COS spectrum of HS\,1914+7139, except for the top-left panel, which is the LAB \hi\ 21-cm emission spectrum toward  HS\,1914+7139. The green colored region denotes the HVC centered at $-180$ \km\ seen solely in absorption while the red colored region is the HVC at $-118$ \km\ seen in both emission and absorption. The dashed vertical lines show the average velocities for the two HVCs. Note that the \oi\ (especially at low velocity) seems affected by terrestrial airglow emission while other saturated absorption lines show that the background subtraction is not always correct yet. \label{f-spectra}}
\end{figure}

The COS observations of HS\,1914+7139 were obtained on October 9 2009 using the gratings G130M (1150--1450 \AA) and G160M (1405--1775 \AA). The exposures times for these two settings were 1.1  and 1.3 ks, respectively, giving S/N\,$\sim 20$ per resolution element. The UV flux of this star is $3$--$5\times 10^{-14}$ erg\,cm$^{-2}$\,s$^{-1}$\,\AA$^{-1}$, which would have required several tens of ks with STIS E140M  to reach similar S/N.  The data were collected in time-tag mode and were processed using the current version of CALCOS (v2.11b) available at the Multi-Mission Archive at STScI (MAST). As the COS mission is still in its infancy, there are many aspects of the data calibration that are not completely optimal. For example, no flat field has been applied to the data and, based on profiles that are known to be fully saturated (see Fig.~\ref{f-spectra}), there seem to be issues with the background correction in some locations (possibly owing to some contamination by scattered light). The post-launch line spread function (LSF) is also more complicated than the pre-launch LSF with the presence of broad non-Gaussian wings, degrading somewhat the resolution of the COS G130M and G185M data  \citep{ghavamian09}. As we use several absorption lines as well as several transitions of a same ion, these adverse factors do not impede the reliable detection of the HVCs and the measurements of their velocities. However, estimates of the column densities are more subject to the LSF uncertainties, background subtraction, and flat fielding, and therefore column densities listed in this paper should be considered as preliminary. 

The stellar continuum was  simple enough to be fitted with low-order Legendre polynomials ($\le 4$) and in Fig.~\ref{f-spectra}, we show some of the resulting normalized profiles. In this figure, we also show the 21-cm emission data from the 36\arcmin\ Leiden-Argentine-Bonn survey  \cite[LAB,][]{kalberla05} in the direction of HS\,1914+7134. The absorption spectra show two high-velocity components at $v_{\rm LSR} = -180$ and  $-118$ \km, but only the latter component is seen in emission. The component at $v_{\rm LSR} = -180 \pm 5$ \km\ has absorption of \cii, \alii, \siii, \siiii, \siiv, \civ, and \oi. All these species (but the weaker line of the \civ\ doublet) are detected at more than the 3$\sigma$ level. The  velocity is the average of  $v_a = \int_{v_1}^{v_2} v \tau_a(v) dv/ \int_{v_1}^{v_2} \tau_a(v) dv$  (where $\tau_a$ is the apparent optical depth) for  all the detected absorption lines (except \cii, \siii\ $\lambda$1193, and \siiii\ that are very strong lines where lower velocity components may contaminate the HVC component). The profiles were integrated from $v_1 \approx -230$ to $v_2 \approx -155$ \km. For the other component, the same species are observed as well as \sii, \feii, \nni, and possibly \nv. Integrating the profiles from $v_1 \approx -155$ \km\ to $v_2\approx -90$ \km\  of \nni\, \alii, \siii\  ($\lambda$$\lambda$1304, 1526), \feii, \civ, \siiv, and \nv\ yields $v_{\rm LSR} = -118 \pm 4$  \km\ for this HVC, within 1$\sigma$ of the \hi\  average emission velocity.

\section{Results and Discussion} 
HS\,1914+7139 is situated at a distance of 14.9 kpc from the Sun and 6 kpc above the Galactic plane.  Three main features of the Milky Way cross its path: Orion Spur at $v_{\rm LSR} \sim  -40$ \km, the Perseus arm at $v_{\rm LSR} \sim  -75$ \km, and the Outer Arm at $v_{\rm LSR} \sim  -90,-120$ \km. Based on the weak interstellar lines (e.g., \sii\ $\lambda$1250, \Niii\ $\lambda$1370), the average LSR velocity of the low velocity component is $-37 \pm 3$ \km, and hence the low velocity component mostly probes some relative nearby gas. 

Using the rotation curve from \citet{clemens85}, a velocity of about $-120$ \km\ occur in gas at Galactocentric distance $R_{\rm G} \sim 22 $ kpc in the direction of HS\,1914+7139 if the gas corotates with the Galaxy. The Galactocentric distance of the star is  $R^\star_{\rm G} = 23.4$ kpc (assuming $R^\sun_{G} = 8.5$ kpc). Hence combining the information from the kinematics and  distance provides support for an origin of the component at $-118$ \km\ from the Outer Arm, and this is therefore the first firm upper limit on the distance of the Outer Arm gas. We, however, note that both velocities and distance are also consistent with those of the nearby complex C ($d = 10 \pm 2.5$ kpc, Thom et al. 2008, and see also Wakker et al. 2007). 

Based on the rotation curve of the Milky Way, the gas moving at $-180$ \km\ corresponds to $R_{\rm G} \sim 120 $ kpc in a corotating disk/halo in the direction of HS\,1914+7139. The limit from the distance of the star evidently rejects this hypothesis, implying that this component is a genuine HVC in the sense that its LSR velocity is inconsistent with a simple model of differential Galactic rotation. This is therefore an highly ionized HVC that is much closer than its corotating distance implies. Its high negative velocity implies it is plunging onto the Milky Way thick disk. 

We already noted above that the sightline to HS\,1914+7139 lies near complex C, but the absolute LSR velocity of the present HVC is substantially larger than those associated with complex C. Searching for other observed sightlines near HS\,1914+7139, we found 5 QSOs/AGNs that are summarized in Table~\ref{t-extra}. All these lines of sight show both absorption at LSR velocities lower and greater than $-150$ \km. For the absorption centered near $-120$ \km, sightlines with typically $b<28\degr$ (for $l\sim 90$--$110\degr$) are related to the Outer Arm while HVCs at $b>30\degr$ are associated with complex C. However, as \citet{tripp03} also discussed, with the current observational evidence it is not clear if complex C and the Outer Arm gas are different entities or have a relationship, except for the \hi\ maps that indicate that the Outer Arm seems to connect more smoothly to the Galactic disk than Complex C \citep[e.g.,][]{wakker01,tripp03}. 

In what follows (except otherwise stated), we focus solely on the HVCs at $v_{\rm LSR} \la -150$ \km\ where little or no \hi\ 21-cm emission has been detected ($N($\hi$) \la 10^{18.7}$ cm$^{-2}$). Preliminary apparent column densities toward HS\,1914+7139 for \oi\ and \siii\ are: $\log N($\siii$) \simeq 13.6$, $\log N($\oi$) \simeq 13.9$, giving $[$\oi/\siii$] = -0.9$ (where $[X/Y]= \log N(X)/N(Y) - \log (X/Y)_\odot $ and the solar abundances are from Lodders et al. 2009), i.e. the gas is dominantly ionized \citep[e.g.,][]{lehner01a}. Using the results from \citet{tripp03}, $[$\oi/\siii$] \simeq -0.53$ toward 3C\,351, again suggesting a large fraction of ionized gas.  Only \ovi\ and \civ\ are detected toward H\,1821+643 \citep{savage95,oegerle00,sembach03,tripp03}, implying that the gas is highly ionized. For the other sightlines listed in Table~\ref{t-extra}, the presence of \siiii\ or/and other higher ions also implies a large amount of ionized gas. So not only do the HVCs listed in Table~\ref{t-extra} and toward HS\,1914+7139 share similar velocities, but they all have low \hi\ column densities and are largely  ionized.  This suggests they trace a common structure.  HS\,1914+7139 places these HVCs at $d<14.9$ kpc and $z<6$ kpc.  While the origin for these HVCs is still uncertain (see below), they cannot be part of the WHIM at very large distances from the Milky Way. 

\citet{tripp03} derived a metallicity for the HVC toward 3C\,351 $[$O/H$]\ga -1.2$ dex solar based on a comparison of \oi\ and \hi. They denominated this HVC the high-velocity ridge (HVR) and argued the HVR could be the leading edge of complex C that is interacting with the lower halo of the Milky Way. This would explain the ionized nature of the HVC, which is expected according to recent hydrodynamical simulations for a HVC moving in a hot halo and approaching the Galactic plane \citep[e.g.,][]{heitsch09}. The distance of HS\,1914+7139 is consistent with this hypothesis, providing some support to this idea. However, several studies suggest a slight gradient in ${\rm O/H}$ from $-0.01$ to about $-0.07$ dex\,kpc$^{-1}$ \citep[e.g.,][]{daflon04,esteban05}, which could accommodate low metallicity at large distances.  There is, however, a large uncertainty in the gradient and how it varies with the galactic longitude.  Based on chemical evolution models and abundance measurements, \citet{cescutti07} found that the lowest $[{\rm O/H}]$ has a mean and a deviation of $-0.2\pm 0.2$ dex at $R{\rm g} > 9.5$ kpc, which does not favor a galactic fountain phenomenon in the outer region of the Milky Way if the overall metallicity of the HVCs is low ($\ll -0.2$ dex solar). 

In summary, these early \hst/COS results have successfully constrained the distance of two multiphase HVCs toward the direction $l = 103\degr$ and $b=+24 \degr$. One is related to the Outer Arm of the Milky Way and the other one traces infalling gas onto the Milky Way. It is too early to conclude if all the highly ionized HVCs are linked to galactic phenomena rather than the Local Group,  but the two detections of highly ionized HVCs in very different regions of the Galaxy (central and outer regions, see Table~\ref{t-sum}) show that at least some of these low $N($\hi$)$ HVCs are near the Milky Way and associated with infalling gas (likely from a Galactic foutain or  accretion of matter condensing from the Galactic corona). We note that so far only  negative high velocities  have been detected in spectra of stars, and it remains to be seen if  highly ionized HVCs with positive velocities are found at $d<10$--20 kpc from the Milky Way. Once all the observations of our program are completed, we should have a sizeable sample that will help us to better understand the origin(s) of the highly ionized HVCs that cover the Milky Way sky.

\acknowledgments

We greatly appreciate funding support from NASA grant HST-GO-11592.01-A from the Space Telescope Science Institute. We thank our contact scientists, David Sahnow and Shelly Meyett, at STScI for all their helps and useful advices. We wish to acknowledge and thank all the NASA personnel, and in particular the astronauts, who made the servicing mission 4 on the \hst\ a complete success.

\clearpage

\begin{deluxetable}{lccccccc}
\tablewidth{0pc}
\tablecaption{Detected Highly Ionized HVCs toward Galactic Stars \label{t-sum}}
\tabletypesize{\footnotesize}
\tablehead{\colhead{Name} & \colhead{$l$}& \colhead{$b$} & \colhead{Sp. Type}& 
\colhead{$d$}& \colhead{$z$} & \colhead{$v^{\rm HVC}_{\rm LSR}$} & \colhead{Ref.}\\
\colhead{} & \colhead{(\degr)}& \colhead{(\degr)} & \colhead{}& \colhead{(kpc)}& \colhead{(kpc)} & \colhead{(\km)} & \colhead{}
}
\startdata
NGC\,5904-ZNG1 &  3.88   & +46.79  & PAGB      &  7.5 &+5.3   & $(-140,-110)$  	&  1    \\
HS\,1914+7139  &  102.99 & +23.91  & B4\,IV    & 14.9 & +6.0  & $(-180,-118)$  	&  2   
\enddata
\tablerefs{(1) Zech et al. (2008); (2) this paper.}
\end{deluxetable}

\begin{deluxetable}{lcccccc}
\tablewidth{0pc}
\tablecaption{Extragalactic Sightlines near HS\,1914+7139 with Ionized HVCs at $v_{\rm LSR} <-150$ \km \label{t-extra}}
\tabletypesize{\footnotesize}
\tablehead{\colhead{Name} & \colhead{$l$}& \colhead{$b$} & \colhead{$\Delta \theta^a$}& \colhead{$[v_{\rm min},v_{\rm max}]^b$}& \colhead{Species$^c$}  & \colhead{Ref.} \\
\colhead{} & \colhead{(\degr)}& \colhead{(\degr)} & \colhead{(\degr)}& \colhead{(\km)} & \colhead{} & \colhead{}
}
\startdata
Q\,1831+731	& 104.04& 27.40	& 3.6	& $-202,-100$	&  \siiii\     			& 1 \\
H\,1821+643	& 94.00	& 27.41	& 8.8	& $-235,-180$	&  \civ, \ovi\     		& 2 \\
HS\,1700+6416	& 94.40	& 36.16 & 12.7	& $-197,-85 $	&  \siiii\     			& 1 \\
3C\,351		& 90.09	& 36.38	& 16.7	& $-230,-150$	&  \siii, \siiii, \siiv, \civ	& 2 \\
Mrk\,876	& 98.27	& 40.38	& 17.0	& $-220,-155$	&  \cii, \Siii, \ovi     	& 3 
\enddata
\tablecomments{$a$: Angular separation from HS\,1914+7134. $b:$ Velocity interval at which high-velocity absorption is seen; note that for the Q\,1831+731 and HS\,1700+6416, the profiles are saturated and it is not possible to separate the Complex C/Outer Arm component from the more negative velocity component. $c$: Known species (from the literature) detected at $v_{\rm LSR} < -150$ \km.}
\tablerefs{(1) \citet{collins09}; (2) \citet{tripp03}, \citet{savage95}, \citet{sembach03}, \citet{oegerle00}, (3) \citet{fox04}, \citet{collins07}.}
\end{deluxetable}

\end{document}